\documentclass[preprint,12pt]{elsarticle}

\usepackage{amssymb,amsmath}
\usepackage{graphicx}
\usepackage{epstopdf}
\usepackage{bm}

\newcommand{\meq}[1]{\begin{equation} #1 \end{equation}}
\newcommand{\bs}[1]{\boldsymbol{#1}}

\newcommand{\ek}{\epsilon_{\mathbf{k}}}
\newcommand{\Ek}{E_{\mathbf{k}}}
\newcommand{\uk}{u_{\mathbf{k}}}
\newcommand{\vk}{v_{\mathbf{k}}}

\begin{document}

\begin{frontmatter}

\title{The Fermi Gases and Superfluids: Short Review of Experiment and Theory
for Condensed Matter Physicists}

\author{K. Levin}
\ead{levin@@control.uchicago.edu}
\address{James Franck Institute\\
The University of Chicago\\
Chicago, 60637, Ill.}

\author{Randall G. Hulet}
\ead{randy@@rice.edu}
\address{Department of Physics and Astronomy\\
Rice University\\
Houston, Texas 77005}

\begin{abstract}
The study of ultracold atomic Fermi gases is a rapidly exploding
subject which is defining new directions in condensed matter
and atomic physics. Quite generally what makes these
gases so important is their remarkable
tunability and controllability. Using a Feshbach
resonance one can tune the attractive two-body interactions
from weak to strong and thereby make a smooth crossover
from a BCS superfluid of Cooper pairs to a Bose-Einstein
condensed superfluid.
Furthermore, one can tune the population of the two spin
states, allowing observation of exotic spin-polarized
superfluids, such as the Fulde Ferrell Larkin Ovchinnikov (FFLO) phase. A wide array of
powerful characterization tools, which often have direct condensed
matter analogues, are available to the experimenter.
In this Chapter, we
present a general review of the status of these Fermi
gases with the aim of communicating
the excitement and great potential of the field.
\end{abstract}


\end{frontmatter}

\tableofcontents

%\maketitle

\section{Introduction}
\label{sec:I}

The Fermi gases and the Fermi superfluids represent a new class of condensed matter ``materials."
Aside from their neutrality and the fact that they appear in confined geometries (traps), they
possess many essential features found in strongly correlated systems. Adding
to the excitement is the fact that these systems are highly tunable. We will see below
that one can dial-in the strengths of interactions (both repulsive and attractive), the size
and geometry of the (optical) lattice, spin polarizations, as well as other features.
The milestones in
this discovery phase were the creation of a degenerate Fermi gas (1999), the
formation of dimers of fermions (2003), Bose-Einstein condensation (BEC)
of these dimers (late 2003) and finally, condensation of fermionic pairs
(2004).
Many challenges were met and surmounted in the process and within a remarkably short period
of time, researchers were able to observe a new form of ``high temperature" superconductivity
(in the sense of large $T_c/E_F$) and to develop a set of tools to characterize this
new state of matter.
The available set of tools is equally remarkable. The experimental complexity of these
ultracold Fermi gases can not be over-stated. It is not possible to use traditional thermometers
to measure temperature, nor attach leads on a sample (in a current-voltage set up)
to measure the superconducting gap. Nevertheless,
the experimental community has devised ways of
doing these analogue condensed matter experiments, as well as the analogue of
photoemission, transport, neutron scattering and ``magnetic field" experiments.
We touch on all of these briefly in this Chapter.

The extensions of these experiments to optical lattices will be discussed in Chapter
5.
But even in studies of trapped gases without lattices,
which are the focus here, there are exciting opportunities for insights into
many physics sub-disciplines.
This is
based on interest in (i) the strong interaction limit, known as the unitary gas (see Chapter 6)
and (ii) the related smooth evolution of superfluidity from fermionic (BCS) to bosonic (BEC). Interest in the first of these has captured the attention of scientists
who also work on
quark-gluon plasmas, as well as in nuclear and astrophysics. The
BCS-BEC crossover has captured the attention of condensed matter physicists who contemplate
implications for high temperature superconductivity.

\subsection{Theory Summary and Overview}

The background theory for this Chapter focuses on fermionic superfluidity and the
unusual ``normal" states which are present above the transition temperature.
The physics
is relatively simple to appreciate.
Fermionic superfluidity is driven by an attractive interaction between fermions
which leads them to pair up and
thereby introduces boson-like degrees of freedom. These bosons, called ``Cooper pairs",
are driven by statistics to condense at low temperatures, and in the process
form a superfluid state. In the simplest case, the ``Bose condensation" is a macroscopic occupation of a many particle
ground state in which the net pair momentum is zero. A formal machinery for implementing
this picture was presented by Bardeen, Cooper and Schrieffer (BCS) and it has worked remarkably
well for addressing conventional superconductors as well as superfluid helium-3.
With the discovery of the high temperature superconductors there was a
re-examination of BCS theory, not so much because it failed in the well studied
superconductors, but because it began to emerge as a very special case of a much
more general theory.

\begin{figure}[tb]
\centerline{\includegraphics{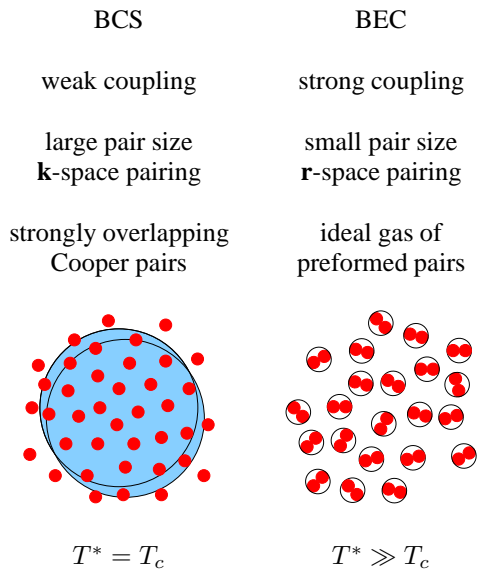}}
 \caption{Contrast between BCS- and BEC-based superfluids}
 \label{fig:crossover}
\end{figure}

This more general theory of fermionic superfluids is known as BCS-Bose Einstein
condensation (BEC) or BCS-BEC theory. This approach
identifies BCS theory with the limit of extremely weak attractive interactions.
This weak attraction is associated with very loosely bound pairs.
We refer to the pair size as the ``coherence length" $\xi$
so that these BCS pairs are
in the limit of very large coherence length
compared to the interparticle spacing. This identification corresponds well
with the behavior of $\xi$ which is directly observed in conventional superconductors.
As the attraction becomes stronger, the pairs become more tightly bound. Since
the superfluid onset temperature, $T_c$, is directly related to this attraction,
it is simultaneously increased.

While there is a smooth crossover between BCS and
BEC superfluids, the physics changes most dramatically when one studies
the behavior above $T_c$. Here, once one leaves the BCS (weak attraction) regime,
the normal phase changes from fermionic to a more bosonic character.
How does one monitor this change in effective statistics? This is possible
through a parameter known as the pairing gap, $\Delta$. The pairing gap
is the energy one must provide to break up the Cooper pairs and create
separate fermions.
We emphasize the pairing gap
parameter in our theoretical discussions.

The cold Fermi gases have provided a unique opportunity to study the BCS-BEC
crossover because
one can continuously tune the attractive interaction via Feshbach resonances.
The literature has focused on the so-called unitary
regime which is roughly mid-way between BCS and BEC. This regime corresponds to
a particular interaction strength in which the two body scattering is
associated with a divergent scattering length $ a \rightarrow \infty$.
One could view this limit as the most strongly interacting regime \cite{Thomas}. 
Deep in
the BCS side the interactions (between fermions) are weak and deep on the BEC
side the interaction (between tightly bound bosons) are similarly weak.
Indeed, many different physics sub-disciplines have been interested in
the unitary gas, which is more extensively discussed in Chapter 6.
One might also imagine that high temperature
cuprate superconductors belong to this more general category of BCS-BEC.
Supporting this scenario is the fact that $\xi$ is anomalously small, and $T_c$
is, of course, high. In addition, there is a rather extensive body of evidence
that the normal state has a non-zero pairing gap.
This is frequently referred to as the ``pseudogap."

\begin{figure}
%\centerline{\Large{ \mbox{\hspace{0.0in} \emph{\small{BCS}} \hspace{0.8in}
%\emph{\small{Unitary}} \hspace{0.8in} \emph{\small{BEC}}} }}
%\vskip 0.05in \centerline{\includegraphics[bb = 0 0 510 150,
%width=3.4in, clip]{cartoon2.eps}}
\centerline{\includegraphics{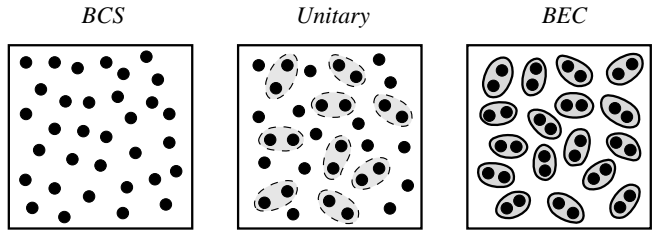}}
\caption{Schematic illustration of excitations (both above and below
the transition $T_c$) in the BCS, unitary and
BEC regimes. The single black discs represent fermionic excitations. Pair
excitations (represented by two fermions)
become progressively dominant as the system evolves from
the BCS to BEC regime.}
\label{fig:Varenna2}
\end{figure}

\begin{figure*}
%\centerline{\includegraphics[width=3.5in,clip]{cartoon1.eps}}
\centerline{\includegraphics[width=3.0in,clip]{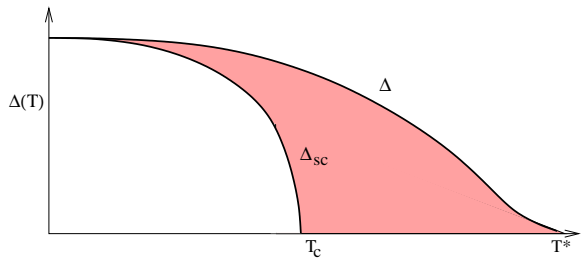}}
\caption{Contrasting behavior of the excitation gap $\Delta(T)$
and superfluid order parameter $\Delta_{sc}(T)$ versus temperature,
appropriate to the unitary regime.
The height of the shaded region roughly reflects the density of
noncondensed pairs at each temperature.}
\label{fig:Varenna1}
\end{figure*}

We begin with an introduction to the qualitative picture
of the BCS-BEC crossover scenario which is represented
schematically in Figure \ref{fig:crossover}. This figure shows the contrasting
behavior of the two endpoints. An important parameter in the literature
is $T^*$ which is the crossover temperature where pairs first start to
form. In the usual BCS theory, the attractive interactions are so
weak that there is no pairing until condensation occurs, so that
$T^* \approx T_c$, while in the BEC limit $T^* >> T_c$.

In Fig.~\ref{fig:Varenna2} we present a schematic plot of the nature of the
excitations (fermionic, bosonic,  or some mix of the two), as one
varies from BCS to BEC. These are present
both above and below $T_c$, the latter as excitations of the condensate.
Midway between
BCS and BEC (i.e., in the unitary regime) there
will be a mix of fermions and quasi-long lived bosons. These bosons and fermions
are not separate fluids, but rather
they are strongly inter-connected. Indeed, the gap in the fermionic
spectrum (related to $\Delta$) is a measure of the number of bosons
in the system.

In Fig.~\ref{fig:Varenna1} we show a schematic of the
gap parameter $\Delta(T)$ as a function of $T$, along with the
superfluid order parameter $\Delta_{sc}(T)$. The former, which
represents the ``bosonic" degrees of freedom, shows that pairs
continuously form once the temperature is less than a crossover temperature
$T^*$, while the order parameter turns on precisely at
$T_c$. The height of the shaded region in this figure reflects the number
of noncondensed pairs. This number increases monotonically with
decreasing $T$, until $T_c$ is reached.  As
$T$ further decreases below
$T_c$ the number of noncondensed pairs begins to decrease monotonically
due to the condensation of zero momentum pairs.

\subsection{Creating Quantum Degenerate Fermi Gases}
\begin{figure}[tb]
\centerline{\includegraphics[width=3.5 in,clip=true,trim=14 14 16 16]{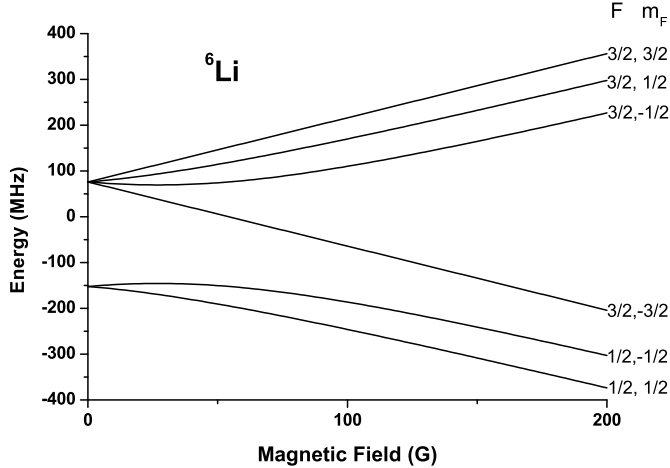}}
\caption{Hyperfine sublevels of the $^6$Li 2S$_{1/2}$ ground state.  The labels on the
the right indicate the total electronic angular momentum $F$ and its projection $m_F$,
which are both good quantum numbers at low fields.  At higher fields, the three most
energetic levels correspond to electron spin-up, while the three lowest to electron
spin-down.  The two lowest-going sublevels, with $m_F = \pm 1/2$, exhibit a broad
Feshbach resonance near 834 G.}
\label{fig:sublevels}
\end{figure}

The achievement of Bose-Einstein condensation of trapped atomic gases in 1995 was a watershed event in the history of many-body physics \cite{Anderson95, Bradley95, Bradley97a, Davis95d}.  Since then, an astounding number of phenomena, described in Chapter 2 of this volume, have been explored with atomic bosons.  Within just a few months of these first experiments, a proposal to use $^6$Li to experimentally realize Cooper pairing in an atomic Fermi gas was published \cite{Houbiers96}. While $s$-wave interactions are forbidden between fermionic atoms in the same internal state due to the Pauli exclusion principle, interactions are
allowed in a two-component Fermi gas.
This
pseudo-spin-1/2 system of atomic fermions can be realized using two of the ground-state hyperfine sublevels, which are shown in Fig.~\ref{fig:sublevels} for $^6$Li.   These sublevels differ in either electronic or nuclear spin projection.  The early proposal showed that the naturally large attractive interaction between the two uppermost (most energetic) sublevels in $^6$Li was sufficient for pairing to occur at temperatures that had already been achieved in the boson experiments.  However, two-body inelastic collisions are unacceptably large for these sublevels.  A better choice is the two \emph{lowest} sublevels of $^6$Li, which are energetically stable.
Furthermore, a Feshbach resonance could be used to tune their relative interaction strength to essentially \emph{any} value \cite{Houbiers98}.
Indeed, the collisional stability of a two-component Fermi system near a Feshbach resonance substantially exceeds that of a Bose gas.
Feshbach resonances have turned out to be essential for experiments on Fermi superfluidity, both in $^{40}$K and $^6$Li; they will be discussed in more detail in the next section.

Creating degenerate Fermi gases proved to be not quite as straightforward as it was for the Bose gases, described in Chapter 1.  The final cooling step in every successful quantum gas experiment with ultracold atoms has been evaporative cooling.  Here, the most energetic atoms are removed from the gas leaving the remaining atoms to rethermalize to a lower temperature.  This process can be very efficient, but it requires these thermalizing collisions to repopulate the high-energy tail of the Boltzmann distribution.  As we've seen above, however, a Fermi gas must contain at least two spin-states (hyperfine sublevels) in order for such collisions to occur, and the additional state opens up more pathways for inelastic loss.  Since all early quantum gas experiments utilized magnetic trapping, a further constraint was that the two sublevels had to be ``weak-field seeking", such that the energy of the state increases with field. Accommodating evaporative cooling without unacceptably high atom loss from these inelastic pathways proved to be considerably more difficult for fermions than for bosons.

\begin{figure}[tb]
\centerline{\includegraphics[width=3.in,clip=true,bb=13 163 600 610, angle=180]{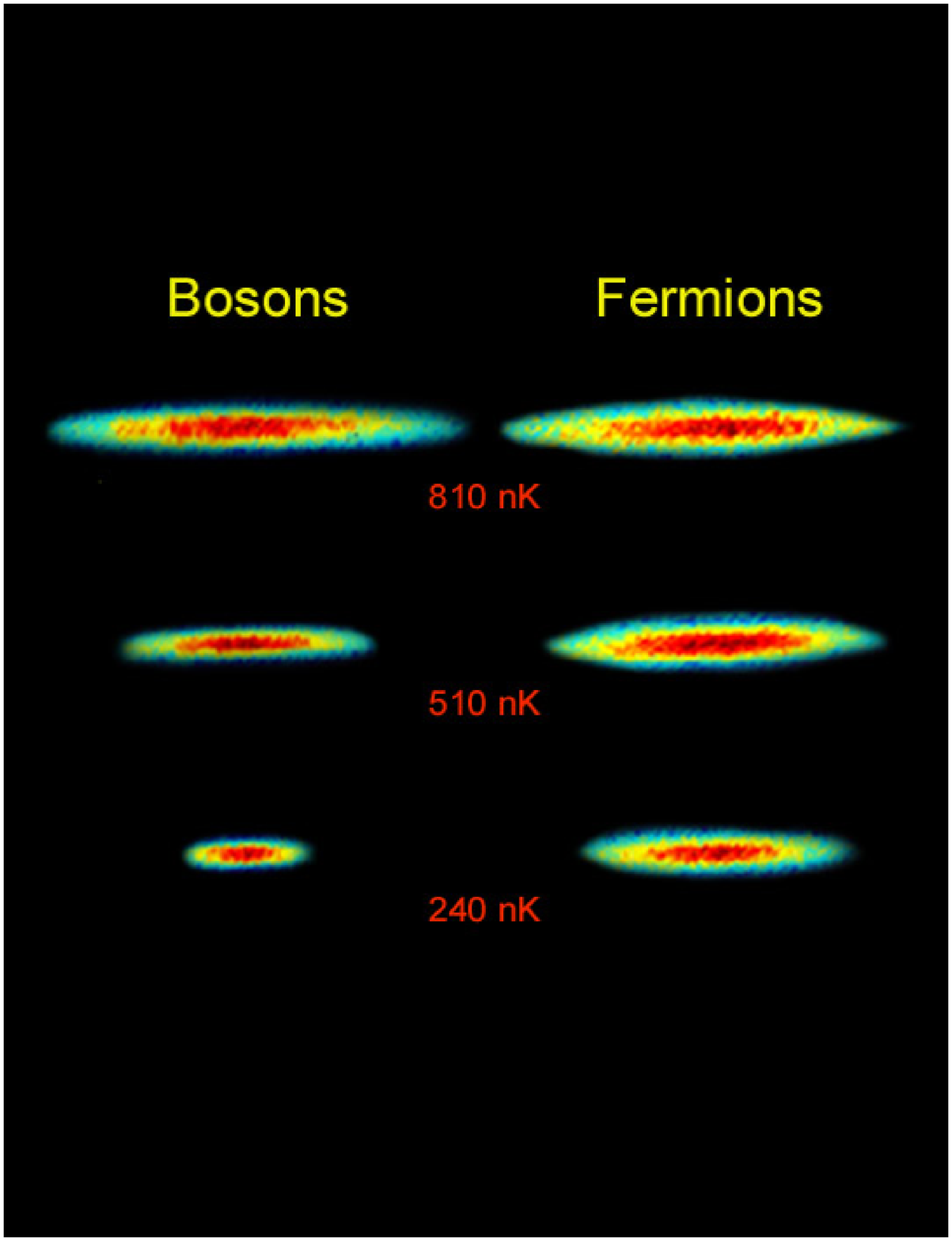}}
\caption{Sympathetic cooling of fermionic $^6$Li by bosonic $^7$Li.  The $^7$Li atoms are actively
evaporated using an RF method, while the $^6$Li is sympathetically cooled via elastic collisions.
The three different temperatures shown, corresponding to the three rows, are obtained by modifying the evaporation cycles appropriately.  A separate laser beam is used to image each isotope. (Data from Ref.~\cite{Truscott01})}
\label{fig:FermiPressure}
\end{figure}

Several methods were developed to circumvent the evaporative cooling problem.  The first degenerate Fermi gas was produced at JILA in 1999 using $^{40}$K \cite{DeMarco99}, which has an unusually large nuclear spin of  4.  Because of the large nuclear spin in $^{40}$K, there are several weak-field seeking sublevels in the lower, more stable manifold, and a gas formed from two of these was magnetically trapped and evaporated to degeneracy.  This approach was not available for $^6$Li since its nuclear spin is only 1. Another approach was developed for $^6$Li in which a single spin-state is cooled ``sympathetically" using a co-trapped but entirely different atomic species.  This was employed at Rice and at ENS in Paris using $^7$Li as the refrigerant atom \cite{Truscott01, Schreck01}.  The actively evaporated $^7$Li cools the $^6$Li by collisions, which are not symmetry forbidden.  Figure \ref{fig:FermiPressure} shows how the \emph{in situ} column densities of the two species evolve as evaporation progresses.  Even though the two isotopes are co-trapped in the same volume their optical transition wavelengths are sufficiently different that the two species can be independently imaged.  Initially, both species are relatively hot and their distributions are essentially the same.  Importantly, as the atoms get colder, it is clear that the fermions occupy a larger volume than do the bosons. This is an effect of Fermi pressure, the same mechanism that stabilizes white dwarf stars against gravitational collapse.

\subsection{Feshbach Resonances}
The Feshbach resonance has played a central role in achieving pairing in ultracold atomic Fermi gases.  No experiment has thus far demonstrated pairing without employing a Feshbach resonance to create a sufficiently strong
attractive interaction.  Furthermore, the realization of the BEC-BCS crossover relies on tuning the interaction strength between atoms from strongly binding in the BEC regime to weakly attractive on the BCS side.  The simplest way to conceptualize such tunability is to imagine that the interaction between two non-identical fermions is a square well with tunable depth $U$.  The interaction can be described by the $s$-wave scattering length $a$.  For very small $U$,
such that the well is unable to support bound states, $a$ is small and negative, corresponding to a small attractive interaction.  This is the BCS regime where the transition temperature
$T_c \sim \exp(-1/k_Fa)$ is exponentially small.  Here, $k_F$ is the Fermi wavevector, and $k_Fa$ is the natural unit of interaction strength.  As $U$ is increased, $a$
remains negative but increases in magnitude until at sufficiently large $U$ the square well finally supports a bound state.  This point corresponds to a scattering resonance where $a$ goes from $-\infty$ to $+\infty$, and is termed the ``unitarity" limit, where the scattering cross section is maximum.  For even larger $U$, $a$ remains positive but diminishes in magnitude.

\begin{figure}[tb]
\centerline{\includegraphics[width=3.0 in,clip=true,trim=0 0 125 110]{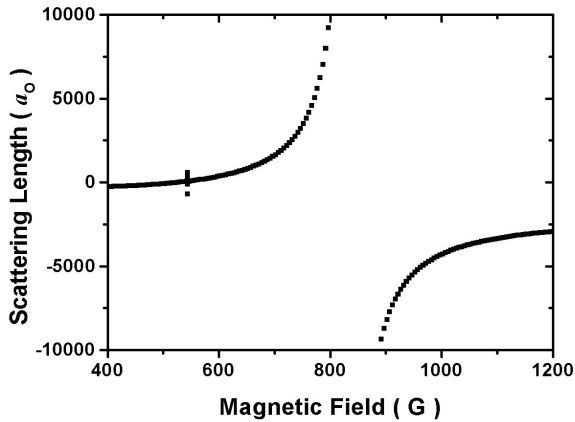}}
\caption{Coupled channels calculations
of the $^6$Li Feshbach resonances involving the energetically two lowest Zeeman sublevels.
A broad resonance is located near 834 G, while a narrow one can be discerned near 543 G \cite{Strecker03}.}
\label{fig:Li6FR}
\end{figure}

While this simple ``single-channel" model gives the flavor of the Feshbach resonance, it is incomplete. The Feshbach resonance actually involves two-channels, usually corresponding to the singlet ($S=0$) and triplet ($S=1$) interaction potentials between pairs of ground-state alkali-metal atoms \cite{ChinRMP10}.  The total electronic spin, $S$, is only an approximately good quantum number, as the singlet and triplet states are coupled by the hyperfine interaction.  A Feshbach resonance occurs when atoms in the ``open" ($S=1$) or scattering channel, are near resonance with a bound state in the ``closed" ($S=0$) channel.  If the open and closed channels have different magnetic moments, the resonance can be tuned magnetically.  A good example is $^6$Li, whose two lowest sublevels (Fig.~\ref{fig:sublevels}) go through resonance near 800 G \cite{Houbiers98}, as shown in Fig.~\ref{fig:Li6FR}.  At a field of 834 G the scattering continuum of the open channel is resonant with the bound singlet.  At higher fields there is no bound state possible, only an attraction that gives rise to pairing in the BCS regime.  But at fields below resonance the superposition of the triplet continuum with the singlet bound state gives rise to a weakly-bound molecular state, whose binding energy scales as $1/a^2$.  Sufficiently far below resonance, the molecular size ($\sim$$a$) will be small compared with the average interparticle distance, giving $k_Fa \ll 1$.  This is the BEC regime where the molecules condense into a Bose superfluid at low temperature.

\section{Establishing Pair Condensation and Superfluidity in Cold Fermi Gases}

One should appreciate that temperature is not straightforward to measure
in these cold gases. Moreover, because they are neutral
and have a normal state gap, it is difficult to convincingly
establish that one has a superfluid phase.
Initial strong indications of pair condensation were first obtained on
the BEC side of resonance \cite{Jin3,Grimm}, where there are
clear bimodal signatures in the density profile.
This bimodality (i.e., separation between condensate and
excited atoms) is the hallmark of Bose superfluids. The density profiles at unitarity, however, have essentially none of this
bimodality.
The earliest
indications of condensation at unitarity came a few months later
\cite{Jin4,Ketterle3}
%,KetterleV,Thomas2,Grimm3,ThermoScience}
via fast magnetic field sweep
experiments which start at unitarity and project onto the BEC regime
(where condensation is more evident in the particle density
profiles).
The presumption is
that even if the condensate fraction is not conserved upon a fast
sweep to BEC, the presence or absence of a condensate will be preserved \cite{Jin4,Ketterle3}.
The time frame for the sweep will not allow a condensate to form in the
BEC regime if there were none present near unitarity, nor will it allow a
condensate to disappear if it was present initially.

\begin{figure}[tb]
%\centerline{\includegraphics[width=2.8in,clip]{LogTemp.eps}}
\centerline{\includegraphics[width=2.8in,clip]{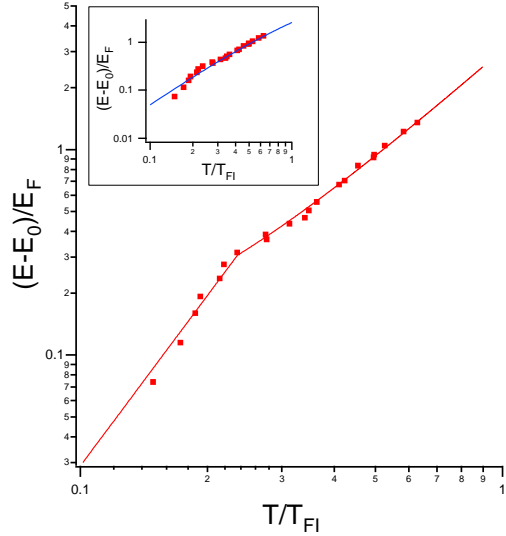}}
\caption{Evidence for a phase transition (presumably to the superfluid phase) via plots of the energy $E$ for a trapped gas vs.~physical temperature $T$. The blue line in the inset
corresponds to the BCS or essentially free Fermi gas case, and the red data points to unitarity.
The slope change of the latter indicates a phase transition.
$T_{FI}$ is the Fermi energy of an ideal Fermi gas, and because these are measurements with a trapped gas,
$T_c \approx 0.23$ differs from the theoretical value obtained for a homogeneous gas of $T_c \approx 0.15-0.17$.
Adapted from Ref.~\cite{ThermoScience}.}
\label{fig:Thomas}
\end{figure}

\begin{figure}[tb]
\centerline{\includegraphics[width=3.0in,clip]{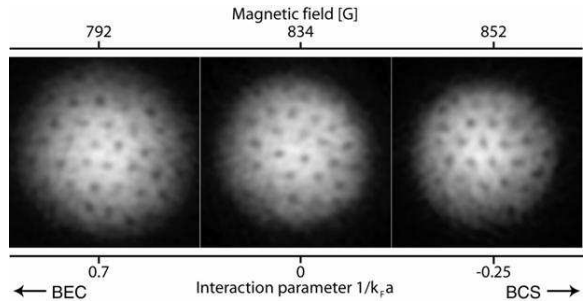}}
\caption{ Evidence \cite{KetterleV} for fermionic superfluidity via
quantized vortices, from BCS to BEC. }
\label{fig:vortices}
\end{figure}

The second generation experiments were based on thermodynamical measurements.  Figure \ref{fig:Thomas} shows evidence for a phase transition as reported in  Ref.~\cite{ThermoScience}.  The measured energy is plotted as a function of temperature.
The key
feature here is that the data (indicated by the solid
circles) show an abrupt
change at a temperature one can call $T_c$. This abrupt change occurs
for the unitary scattering case. No such feature is seen for the
noninteracting Fermi gas.

The last generation experiment to make the case for superfluidity was
the rather stunning observation of quantized vortices by the MIT group
\cite{KetterleV}, which is shown in Fig.~\ref{fig:vortices}. Although, these also involve sweeps to the BEC regime, to obtain
sufficient contrast in the images, they provide the most direct evidence for
superfluidity in the unitary gases.

\section{Theory Outline}

In contrast to the Bose gases, there is no consensus theory as yet to describe these
Fermi superfluids. Considerable effort has gone into
both many body analytic schemes as well as Monte Carlo and related numerical approaches.
Measures of what constitutes a ``successful" theory differ from one sub-community to another.
From a condensed matter perspective one tends to look for novel physical effects and focus on conceptual
issues relating, for example to other highly correlated systems. Nevertheless, the universality
which appears precisely at unitarity, and the history of precision measurements associated with the atomic physics community tend to favor theoretical schemes
which make quantitative contact with experiment.

In this Chapter we restrict our consideration
to analytical studies which build on the simplest (BCS) theory of
conventional superconductors. This is largely because BCS theory represents, perhaps,
the most complete and accessible analytical theory we have in condensed
matter physics. We will see below that the form of the ground state
wavefunction associated with the BCS-BEC crossover is the same as that
introduced in the original BCS theory. One has the hope, then, of
being able to construct  a theory of the crossover in as complete
a fashion as the 1957 theory of Bardeen, Cooper and Schrieffer.
This is not to say that the ultimate theory of the unitary gas is
expected to be fully captured by this mean field approach, but it nevertheless
should provide some useful intuition.

The field of BCS-BEC crossover is built around early observations by
Eagles \cite{Eagles} and Leggett \cite{Leggett} that the BCS ground
state is much more
general than was originally believed.  If one increases the strength of
the attraction and self-consistently solves for the fermionic chemical
potential $\mu$ (which eventually decreases from the
Fermi energy to negative values), the wave function corresponds to a more BEC-like
form of superfluidity.
This ground state
is given by the standard
BCS wavefunction,
\begin{equation}
\Psi_0=\Pi_{\bf k}(\uk+\vk c_{\bf k,\uparrow}^{\dagger}
c_{\bf -k,\downarrow}^{\dagger})|0\rangle \,,
\label{eq:1a}
\end{equation}
where $ c^\dag_{\bf k,\sigma}$ and $ c^{}_{\bf k,\sigma}$ are the
creation and annihilation operators for fermions of momentum ${\bf k}$
and spin $\sigma=\uparrow,\downarrow$.  The variational parameters $\vk$
and $u_{\bf k}$ are associated with the number of occupied and
unoccupied pair states, respectively.

\subsection{Theory of Finite Temperature Effects:
Comparing BCS and Ideal Boson BEC}

It is useful to base our intuition of fermionic
superfluids on that of true Bose gas condensation.
The bosonic degrees of freedom appearing in the BCS wavefunction
Eq. (\ref{eq:1a}) are associated with fermionic pairs. Pair-pair
interactions are not explicitly present so one could say these
Cooper pair bosons
are essentially ideal.
Since an ideal Bose gas cannot support superfluidity, the
superfluidity in the BCS case (as established, say, by
the Meissner effect)
implies that
fermionic substructure of the bosons is necessary and sufficient to
sustain superfluidity.
For ideal point bosons one uses the condition
that the total number of bosons is the sum of the condensed contribution
$N_0(T)$ and the excited contribution $N'(T)$ with
$N = N_0(T) + N'(T)$. Moreover, the latter is straightforward to calculate
\begin{equation}
N'(T)=\sum_{\mathbf{q}\neq
      0}b(\Omega_{q})
\end{equation}
where
$b(\Omega_{q})$ is the usual Bose-Einstein function written in terms of
the non-condensed bosonic excitation spectrum $\Omega_q \propto q^2$.
Converting this to
a density of states integral one has,
following Chapter 2,
\begin{equation}\label{N'}
N'(T) = \int_{0 }^\infty d\epsilon\,\frac{g(\epsilon)}{\exp[\beta(\epsilon-\mu_{B} )]-1}
\end{equation}
which defines the total number of particles not in the condensate.
Here $g(\epsilon)$ is the density of states per unit volume for bosons with
$\epsilon \propto q^2$ dispersion and $\epsilon_0 $ is the
single particle ground state energy which we take to be zero, as in Chapter 2.
The equality $\mu_{B}(T,N)=0$
is a fundamental constraint for all $ T \leq T_c$.
The condensate fraction $N_0$ is obtained
from $N - N'(T)$.

To arrive at the fermionic counterpart of the above equations,
we need to formulate a generalized many body theory.
A number of such schemes have been introduced
in the literature
but we begin with one which is consistent with
Eq. (\ref{eq:1a}) and with finite temperature
extensions associated with Gor'kov and Bogoliubov-de Gennes theory.
The simplest extension of BCS theory to address the crossover can be summarized via the following
equations. As with the ideal Bose gas one implements a constraint,
on the non-condensed fermion
pairs which, below $T_c$  are in chemical equilibrium with the condensate
\begin{equation}
\mu_{pair} = 0 , \qquad T \le T_c \;.
\label{eq:21}
\end{equation}
Importantly, if the non-condensed pairs are properly identified,
this leads to the familiar BCS equation for the
pairing gap
\begin{equation}
\Delta(T) =-U \sum_{\bf k} \Delta(T)
 \frac{1-2 f(\Ek)}{2 \Ek}\,, \\
\label{eq:19}
\end{equation}
where $\Ek = \sqrt{ (\ek -\mu)^2 + \Delta^2 (T) }$, and $f(\Ek)$ is the usual
Fermi function.
We may decompose the excitation gap into two contributions
\begin{equation}
\Delta^2 (T)  = \Delta_{sc}^2 (T) + \Delta_{pg}^2 (T),
\label{eq:13}
\end{equation}
where $\Delta_{sc}(T)$ corresponds to condensed and $\Delta_{pg}(T)$ to
the non-condensed gap component.
Just as in the Bose case, the
number of non-condensed bosons is determined from the dispersion
of the non-condensed pairs which must be compatible with
Eq. (\ref{eq:19}) and yields
\begin{equation}
\Delta_{pg}^2 (T) = Z^{-1}\sum b(\Omega_q, T) \,.
\label{eq:81}
\end{equation}
Here $Z$ is a coefficient of proportionality (unimportant for our
purposes) which can also be
determined microscopically.
It should be stressed that
while $\Delta^2(T)$ plays a similar role in the fermionic system to
the total bosonic particle number $N$, the former is generally temperature
dependent (except in the extreme BEC limit.)
Finally, not only does one need the bosonic chemical
potential (for bosons in equilibrium with the condensate),
but the fermionic chemical
potential of the excited fermions must be established in the usual way via
the well known number equation.
\begin{equation}
n  = \sum _{\bf k} \left[ 1 -\frac{\ek - \mu}{\Ek}
+2\frac{\ek - \mu}{\Ek}f(\Ek)  \right] \,,
\label{eq:19b}
\end{equation}
In this way, as follows from Eq. (\ref{eq:13}),
the onset of superfluid coherence or non-zero $\Delta_{sc}$ is
associated with the condition that the gap equation cannot
be satisfied by having only non-condensed pairs.
We emphasize that
$\Delta^2(T)$ plays an analogous role to the
number of bosons $N$, and this should underline the fact noted
earlier that $\Delta \neq 0$ is really the only way to get
a handle on the bosonic degrees of freedom in a fermionic system.

\subsection{Analytic Formalisms for Addressing Fermi Gas Experiments}

What is particularly distinctive about the Fermi gases as compared to their
Bose counterparts is the fact that the former can be studied throughout the
entire range of temperatures and, moreover, one finds the expected
second order phase transition at $T_c$. For the Bose gases, theories
are mostly confined to very low $T$ and, when extended, lead to a first
order phase transition at $T_c$.
In Chapter 2 we saw that a well developed tool for addressing experiments
in the cold Bose gases was the Gross-Pitaevskii (GP) theory which could
be used in both time independent and time dependent situations.
In order to apply the theoretical ideas discussed above to experimental situations
in the Fermi gases, where there are spatial dependencies, one has
three main analytical tools.
(i)
Landau-Ginsburg theory which is the fermionic
analogue of Gross-Pitaevskii theory which
describes the condensate and (ii) Gor'kov theory and (iii) Bogoliubov-de Gennes theory.
The last two describe the fermionic excitations of the condensate but become more complicated when one includes non-condensed bosons.
  An additional analytical tool which is of widespread utility is linear response
theory to address weak perturbations of the superfluid.
An analytical many body approach to BCS-BEC schemes lends itself to
implementation of linear response theory, provided one does this in
a consistent, conservation-law-respecting way.
The larger number of theoretical options for the Fermi gases as compared with
their Bose counterparts reflects the fact that they contain three rather than
two components: fermionic excitations, pair excitations and condensate contributions.
In strict BCS theory the situation is simpler since there are no pair excitations, while
in strict BEC theory there are no fermionic excitations.

We saw in Chapter 2 that the (Bose gas) \textit{condensate} dynamics is given by
\begin{equation}\label{TDGP}
i\hbar\frac{\partial \Psi_{B}(\bm r,t)}{\partial t} = \left[-\frac{\hbar^2\nabla^2}{2M} + V_{\rm tr}(\bm r) + g|\Psi_{B}(\bm r,t)|^2\right]\Psi_{B}(\bm r,t),
\end{equation}
where $\Psi_{B}$ now depends on $t$ as well as on $\bm r$.
A similar equation can be written for Fermi gas, although
here one should be careful to address both
the dynamics of the condensate as well as that of
the non-condensed pairs. For both of these, one writes rather generally
the same equation as for the bosons, but with a different pre-factor on the left
\begin{equation}\label{TDGP-b}
(e ^{i \theta}) \times \hbar\frac{\partial \Psi_{F}(\bm r,t)}{\partial t} = \left[-\frac{\hbar^2\nabla^2}{2M} + V_{\rm tr}(\bm r) + g|\Psi_{F}(\bm r,t)|^2\right]\Psi_{F}(\bm r,t),
\end{equation}
This equation is known
as the time-dependent Landau-Ginsburg theory, which has been derived
microscopically for the condensate only near $T_c$ and in the BCS regime.
Here one finds a diffusive dynamics in contrast to the behavior of
the time dependent bosonic Gross Pitaevskii behavior, albeit primarily
associated with very low temperatures.
Thus, the factor $ e ^{i \theta}$
is purely real for the fermionic condensate in this temperature
regime.
Here the bosons are not well established
or long-lived.
Deep in the BEC regime of fermionic superfluids
the dynamics associated with the non-condensed pairs is such
that $ e ^{i \theta}$
is purely imaginary, corresponding to stable, long lived
bosons which
have a propagating dynamics. More generally, between
BCS and BEC this factor is a complex number and
$\theta$ may be viewed as varying with the strength of the attractive
interaction.

By far the most straightforward way of including trap inhomogeneity
effects
is the local density
approximation (LDA). This approximation assigns the properties of a
non-uniform fermionic system at a given point their bulk values with an effective
local chemical potential.
Then the calculations
proceed as in a homogeneous system with the replacement
\begin{equation}
\mu(\bm {r}) = \mu_o - V_{\rm tr} (\bm {r}),
\end{equation}
where $ V_{\rm tr} (\bm {r})$
represents the confining potential.  Here the fermionic chemical
potential $\mu(\bm{r})$ can be viewed as varying locally
but self consistently throughout the trap and $\mu_o$ is the chemical
potential at the trap center.
For the most part, this approach has been useful for addressing
thermodynamic properties in a trap.

\section{Experimental Tools}

A remarkable series of advances have made it possible to find experimental
counterparts to many of the most powerful tools we have in condensed matter
physics. These are outlined in Table 1 below.
In the next few subsections we discuss how these are implemented
and some of the observations based
on these techniques.

\vskip3mm
\begin{table}
\begin{tabular}{|p{1.35in}|p{2.05in}|p{1.3in}|}
\hline
& \parbox[c][10mm][c]{2.in}{\centering{Cold Fermi gases}} &
\parbox[c][10mm][c]{1.3in}{\centering{Condensed Matter}} \\
\hline
\parbox[c][10mm][c]{1.3in}{\centering{Gap Measurements}} &
\parbox[c][10mm][c]{2.in}{\centering{Radio Frequency Spectroscopy}}&
\parbox[c][10mm][c]{1.3in}{\centering{Tunneling Measurements}} \\
\hline
\parbox[c][10mm][c]{1.3in}{\centering{Fermionic Dispersion}} &
\parbox[c][10mm][c]{2.in}
{\centering{Momentum Resolved Radio Frequency}}
& \parbox[c][10mm][c]{1.3in}{\centering{Angle Resolved Photoemission}} \\
\hline
\parbox[c][10mm][c]{1.3in}{\centering{Scattering Measurements}} &
\parbox[c][10mm][c]{2.in}{\centering{Bragg two-photon scattering}}&
\parbox[c][10mm][c]{1.3in}{\centering{Neutron Scattering}}\\
\hline
\parbox[c][10mm][c]{1.3in}{\centering{Transport Measurements}} &
\parbox[c][10mm][c]{2.in}{\centering{Viscosity and Spin Transport}}&
\parbox[c][10mm][c]{1.3in}{\centering{Conductivity dc and ac}}\\
\hline
\parbox[c][10mm][c]{1.3in}{\centering{Magnetic Field Studies}} &
\parbox[c][10mm][c]{2.in}{\centering{Critical Rotation Frequency}}&
\parbox[c][10mm][c]{1.3in}{\centering{Upper Critical Magnetic Field}}\\
\hline
\end{tabular}
\caption{Summary of the analogous experimental
probes used in trapped atomic gases and their counterparts in condensed matter.}
\end{table}

\subsection{Measuring the Pairing Gap}

Experiment and theory have worked hand in hand in developing an
understanding of the so-called radio frequency (RF) ``pairing gap spectroscopy'' in the
atomic Fermi gases. This class of experiments was originally suggested
as a method
for establishing the presence of superfluidity \cite{Torma,Torma4}.
Pairing gap spectroscopy is based on using a third
atomic level, called $|3 \rangle$, which does not participate in the
superfluid pairing. Under application of RF fields, one component of the
Cooper pairs, called $|2 \rangle$, is excited to state $|3\rangle$.  If
there is no gap $\Delta$ then the energy it takes to excite $|2 \rangle$
to $|3 \rangle$ is the atomic level splitting $\omega_{23}$. In the
presence of pairing (either above or below $T_c$) an extra energy
$\Delta$ must be input to excite the state $|2 \rangle$, as a result of
the breaking of the pairs.

\begin{figure}[thb]
\centerline{\includegraphics[width=2.5in,height=2.5in,clip]
{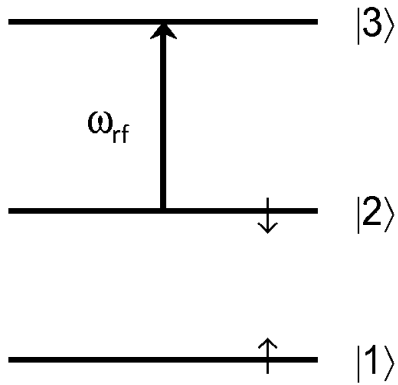}}
%{YanHe_page7.eps}}
\caption{Experimental configuration for radio frequency spectroscopy.}
\label{fig:rf}
\end{figure}

%\begin{figure}[tb]
%\centerline{\includegraphics[
%bb=218 248 391 435,
%width=2in,angle=270,clip]
%{rfspect.eps}}
%\caption{Experimental configuration for radio frequency spectroscopy.}
%\label{fig:rf}
%\end{figure}

A ground breaking experimental paper \cite{Grimm4} reported the first
experimental implementation of this pairing gap spectroscopy in $^6$Li
over a range of fields corresponding to the BCS, BEC and unitary
regimes. Accompanying this paper was a theoretical study \cite{Torma2}
based on the BCS-BEC crossover approach
introduced earlier \cite{JS2}, but, importantly, generalized to include
trap effects.
Indeed, because of trap effects measurements of the pairing
gap are not entirely straightforward to interpret. At general
temperatures, a measurement of the current out of state $|2 \rangle$
is associated with
two discrete structures.
A sharp
peak at
$\omega_{23} \equiv 0$ which derives from ``free" fermions at the
trap edge and a broader peak which reflects the presence of paired
atoms; more precisely, this broad peak corresponds to the distribution of
$\Delta$ in the trap.  At high $T$ (compared to $\Delta$), only the
sharp feature is present, whereas at low $T$ only the broad feature
remains.
Additional experiments have
introduced a powerful way of exploiting and enhancing RF spectroscopy using tomographic techniques
\cite{MITtomo}.
Here the RF contribution is resolved at different distances
from the trap center, throughout the trap. This spatial distribution
is obtained using in-situ phase-contrast imaging and 3D image
reconstruction.
In this way,
scans at different trap radii yield an
effectively homogeneous spectrum.

These data alone do
not directly indicate the presence of superfluidity, but rather they
provide evidence for pairing.
Indeed, like photoemission in condensed matter systems, these measurements reflect the
fermionic  spectral function $A({\bf k}, \omega)$.
One caveat should be noted here. Unlike a photoemission experiment where the fermion
is removed from the sample, here it is excited to a higher internal energy state.
As a consequence there may be residual interactions between atoms in this excited
state and the non-excited (``Cooper pair partner") states in the system.
These are known as final state effects, which can, fortunately, often
be minimized.

\subsection{Momentum Resolved Radio Frequency
Experiments: A Cold Gas Analogue of Angle Resolved Photoemission}

Recent experiments on $^{40}$K from the JILA group \cite{Jin6}
have demonstrated that it is possible to measure spectral
functions directly using momentum resolved RF pairing gap spectroscopy
over a range of magnetic fields throughout the BCS-BEC crossover. These
experiments are able to resolve the kinetic energy $E_k$, and
thereby, the three-dimensional momentum distribution of the RF-excited
(or ``out-coupled'') state 3 atoms. Since the momentum of the RF photon
is effectively negligible, the momenta of the out-coupled atoms can
be used to deduce that of the original 1-2 paired state.  There is a
substantial advantage of using $^{40}$K for these studies over the more widely studied
$^6$Li since there are no nearby Feshbach resonances involving the final state for $^{40}$K that complicate interpretation of the spectra.
Momentum resolved RF spectra can be compared with
momentum resolved (or ``angle resolved'') photoemission in the high
temperature superconductors.
The goal of these experiments and related theory is to deduce the
fermionic quasi-particle dispersion, which would
reveal the pairing gap $\Delta(T)$.

\begin{figure}[tb]
\begin{center}
\includegraphics[width=1.5in,clip]
{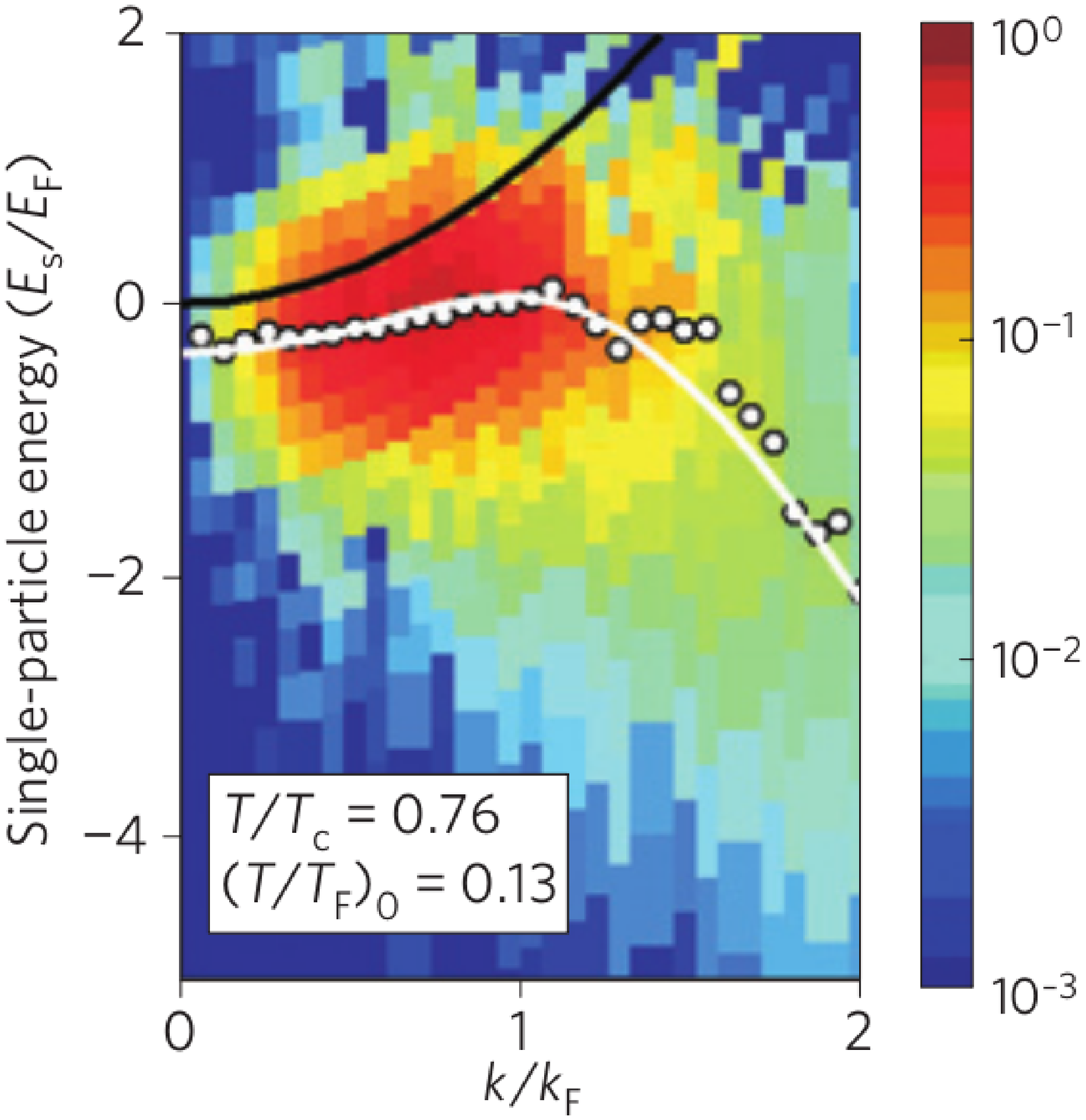}
\includegraphics[width=1.5in,clip]
{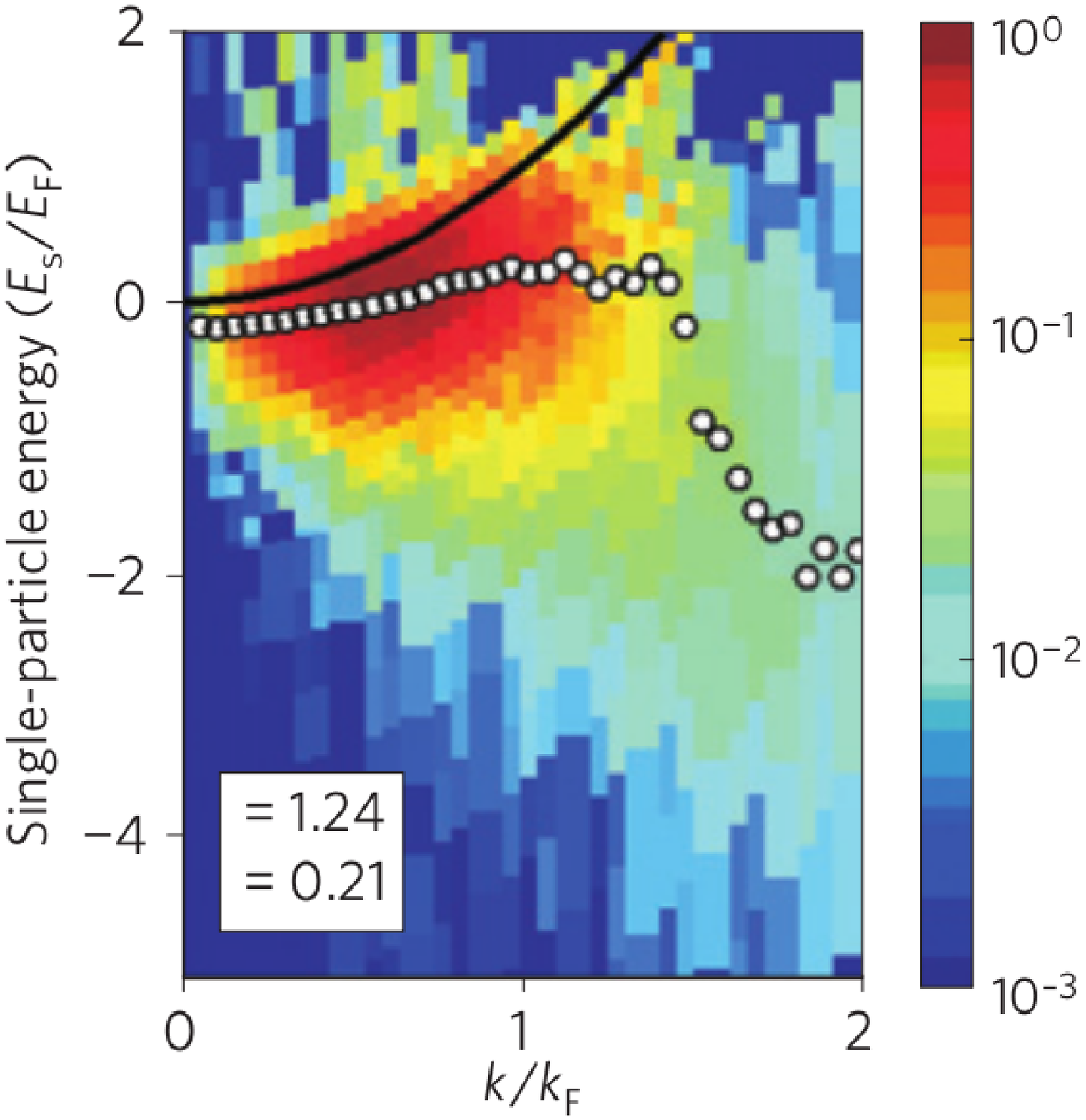}
\includegraphics[width=1.5in,clip]
{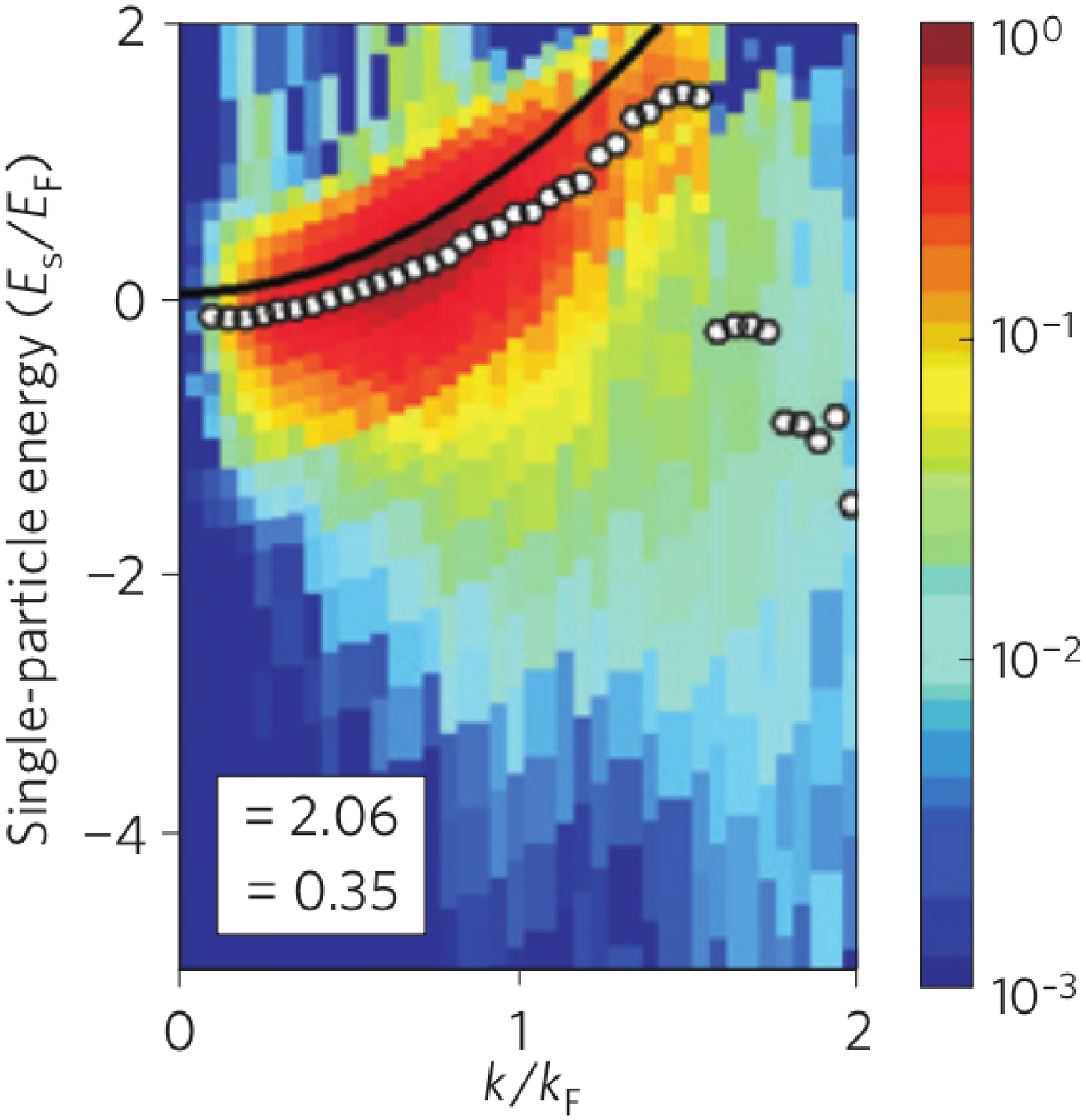}
\caption{Momentum resolved radio frequency (``Photoemission")
spectra \cite{Jin7} throughout the pseudogap regime. These spectra are
for Fermi gases at three different temperatures, each with roughly
the same interaction strength, near unitarity.}
\label{fig:Jin}
\end{center}
\end{figure}

In Fig.~\ref{fig:Jin} we present
experimental measurements of the one particle
fermionic spectral functions as a contour plot.
The
dotted white curve represents an estimate of the experimentally
deduced peak dispersion, which can then
be fit to the BCS
dispersion involving $\Ek$, which was introduced below
Eq.~(\ref{eq:19}).
With
higher
resolution it should be possible to obtain more direct information
about the mean experimentally-deduced gap size.
Importantly, the fact that
the experiments were done near $T_c$
has been argued to suggest that there is a sizable pseudogap in the Fermi
gases at and above $T_c$ in the unitary regime.

\subsection{Universal Properties: The Closed Channel Fraction and
the Contact}

Strongly interacting Fermi gases exhibit a universality, as discussed in Chapter 6, that extends their significance beyond that of any particular realization, such as cold atoms, atomic nuclei, or even quark matter.  At unitarity, the absence of any length scale other than the particle density implies that there is a direct proportionality
between the chemical potential and the Fermi energy of a non-interacting
gas. The coefficient of proportionality, known as the Bertsch parameter, governs the low energy properties of the system.   Short distance correlations, characterized by a parameter $\mathcal{C}$, known as the ``contact", have also been shown to be related to a broad array of universal properties \cite{TanAnnPhys1,TanAnnPhys2, Braaten08}.  $\mathcal{C}$ connects quantities as diverse as the high-momentum tail of the momentum distribution (including the high frequency tail of RF spectra), the total energy, the rate of change of total energy with respect to an adiabatic change in $a$, and a virial theorem relation \cite{TanAnnPhys1,TanAnnPhys2, Braaten08}.  These universal relations are particularly powerful as they extend beyond unitarity into the BCS and BEC regimes, connect microscopic quantities to thermodynamic ones, and require only that $|a|$ be large compared to the scale of the interaction potential.

$\mathcal{C}$ can be directly measured by determining the strength of the local pair correlations, which was done using photoassociation \cite{Partridge05}.  We saw in section 1.3 that Cooper pairs in these systems are a superposition of the triplet scattering state (open channel) and a bound vibrational level of the singlet potential (closed channel).  For $^6$Li, the bound $S=0$ vibrational level corresponds to $v=38$, and the pairs can be expressed as
\begin{equation}
|\psi_{\rm p}\rangle=Z^{1/2}|\psi_{v=\,38}(S=0)\rangle +
(1-Z)^{1/2}|\phi_{\rm a}(S=1)\rangle, \label{eq:dressed_mol}
\end{equation}
where $Z$ is the closed-channel fraction.  Here, $\psi_{v=\,38}(S=0)$ are the closed-channel molecules and $\phi_{\rm a}(S=1)$ are the free atom pairs in the triplet channel.  In the case of wide resonances, and the $^6$Li resonance is as wide as any known Feshbach resonance, $Z$ is expected to be small throughout the resonance region.  For sufficiently small $Z$, the resonance may be well described by a universal single-channel model, such as the square-well discussed previously.  Under these conditions, the macroscopic properties of the superfluid are independent of the microscopic physics of the two-body interactions.

\begin{figure}[tb]
\centerline{\includegraphics[scale=0.35, bb=50 192 689 675]{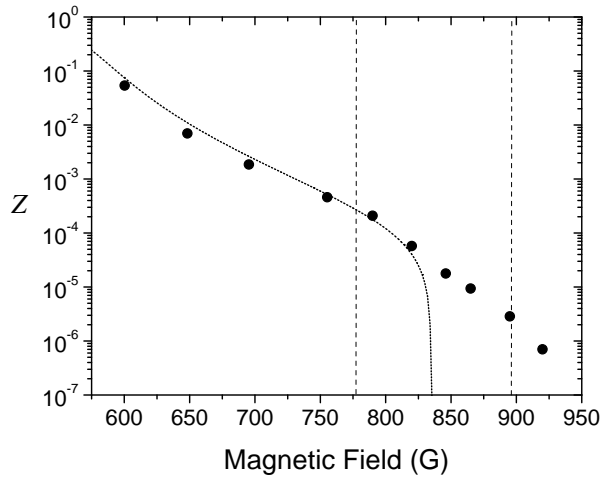}}
\caption{Closed-channel fraction $Z$ throughout the BEC-BCS crossover.
The closed circles represent the value of $Z$ extracted from the rates
of photo-excitation to an excited molecular level.
The dotted line shows a comparison with results obtained from an exact
(coupled channels) two-body calculation. The vertical dashed lines
represent the boundaries of the strongly-interacting regime, $k_F|a|
> 1$.  (Reprinted from Ref.~\cite{Partridge05})}.
\label{fig:ZvsB}
\end{figure}

The quantity $Z$ has been measured experimentally for $^6$Li using photoassociation \cite{Partridge05}.  Since $S$ is a good quantum number in the vibrational levels of the Li$_2$ molecule, the selection rule $\Delta S=0$ is obeyed in the photoassociation transition.  The singlet part of the pairs can then be picked out by driving an optical transition from $|\psi_{\rm p}\rangle$ to an electronically-excited molecular state with $S=0$. The rate of such an excitation will be proportional to $Z$, where the proportionality depends on the constant bound-bound matrix element between $|\psi_{v=\,38}(S=0)\rangle$ and the excited molecule.  An excitation results in a detectable loss of trapped atoms.  The rate of excitation was measured in this way and the corresponding values of $Z$ were determined throughout the BEC-BCS crossover, as shown in Fig.~\ref{fig:ZvsB}.  At unitarity, $Z < 10^{-4}$, and $Z$ remains smaller than 1\%, even deep into the BEC regime.  These results confirm the universality for broad resonances.

It was pointed out in Ref.~\cite{Partridge05} that since the rate of photo-excitation in this experiment is proportional to the overlap between two tightly bound molecular levels of the Li$_2$ molecule whose sizes are much smaller than typical interparticle distances, it is also proportional to the integral over volume of the local pair correlation $G_{2}(r,r)=\langle\hat{\psi}^{\dagger}_{\downarrow}(r)\hat{\psi}^{\dagger}_{\uparrow}(r)
\hat{\psi}_{\uparrow}(r)\hat{\psi}_{\downarrow}(r) \rangle$, where
$\hat{\psi}_{\uparrow}$ and $\hat{\psi}_{\downarrow}$ are the
fermionic field operators for atoms in different internal states.  Consequently, a measurement of $Z$ also corresponds to a measurement of the short-range pair correlations.  Furthermore, since the integral over volume of $G_{2}(r,r)$ is proportional to the contact $\mathcal{C}$ introduced by
Tan in Refs.~\cite{TanAnnPhys1, TanAnnPhys2}, the measurement of the closed-channel fraction is a measurement of the contact \cite{Werner09, Zhang09}.

Several of the other contact relations have been recently experimentally verified \cite{Stewart10}.  In these measurements, $\mathcal{C}$ was obtained from measurements as diverse as the high-momentum tail of the momentum distribution obtained by releasing the atoms from a trap, to thermodynamic quantities such as a generalized virial relation.

\subsection{Thermodynamics}
The thermodynamic variables, energy, pressure, and entropy, have been
systematically studied for the unitary gases. The earliest such measurements established
trap averaged quantities \cite{ThermoScience}, such as plotted in Fig. 7.  These measurements have been reanalyzed using a new determination of
temperature, independent of
theory. More recently \cite{SalomonFL}, there has been considerable progress
in establishing the equation of state or the thermodynamic potential, $\Omega = - PV$, for a homogeneous gas.
This is based on the local density approximation and
a simple relation between the local pressure inside a trapped gas and the
twice integrated density profiles, or ``axial density".
Here temperature is usually determined by using the surface density as a thermometer.
A single image gives the
pressure as a function of variable chemical potential, thereby providing a
large number of independent determinations of the equation of state.  By collecting and averaging the data from many such images, one obtains the equation of state with very low noise.
Figure 12 shows a comparison of recent data from three different
groups as well as examples of theoretical plots in the inset.
This figure should make it clear that thermodynamic studies (there
are counterparts for entropy, chemical potential, etc.) have
received considerable attention in the cold atom community.

\begin{figure}
\begin{center}
\includegraphics[width=2.80in,clip]
{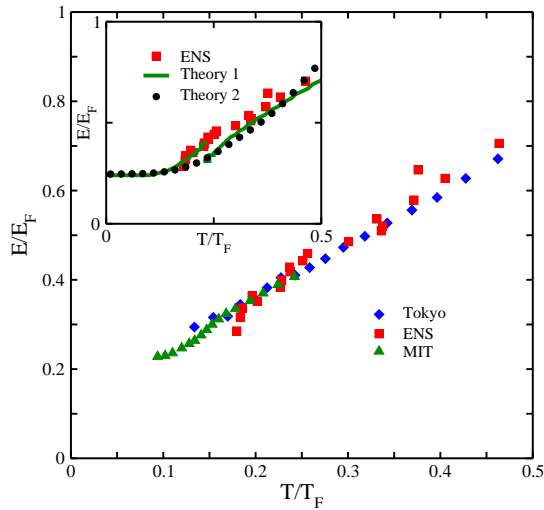}
%{Fig12final.eps}
%{Fig12new.eps}
%{Comparison.eps}
\caption{Comparison of different equations of state for the
energy $E$ versus temperature
measured experimentally and labeled ENS \cite{SalomonFL}, Tokyo
\cite{Mukaiyama} and MIT \cite{ZwierleinThermo}.
The inset presents predictions from two theory groups denoted
1 \cite{Drummond3} and 2 \cite{ThermoScience}, respectively,
as compared with one set of experimental data \cite{SalomonFL}. The
equations of state in some instances have been updated with
new thermometry. The Theory ``2" plots
include a fitted Hartree term and Theory ``1" is based on a modified Nozieres Schmitt-Rink approach \cite{NSR}.}
\label{SEP}
\end{center}
\end{figure}

Of considerable interest in these thermodynamic experiments are universal
properties associated with unitarity. In this infinite scattering length
case (with an interaction of zero range) the energy per unit volume
of the system is directly proportional to that of the free Fermi
gas at the same density $n$.
\begin{equation}
 \epsilon(n) \equiv \frac{E}{V} \propto \frac{n^{5/3}}{m} \equiv \xi 
\epsilon_{free}(n) 
\end{equation}
Here $\xi$ is the ``Bertsch" parameter which is independent
of any materials parameters, applying equally to all Fermi systems
at unitarity.
This parameter appears to be around $0.38$ within about 2 \%.

Experiments are close to converging on these thermodynamical characterizations of
both the trapped and homogeneous gases. The equations of state for the latter,
in particular, have been viewed as important benchmarks for assessing
numerical and analytical approaches to the Fermi gases. Nevertheless it should be stressed that
thermodynamical probes are not as discriminating tests of
theory as are dynamical probes. This is most readily
seen by comparing transport properties (say, the shear viscosities) of fermionic vs
bosonic quantum fluids, helium-3 and helium-4 \cite{NJOP}, with their
specific heat counterparts. Here one sees that the latter are far more similar than
are the viscosities. 
Using thermodynamics as a precision test
of theory may be incomplete and
it is extremely important to provide a characterization of transport in
the ultracold Fermi gases, as well.
This lays the groundwork for
the focus on the dynamics in unitary gases which is discussed in Chapter 6.

\subsection{Transport Experiments in the Fermi Gases}

We summarize briefly these
viscosity measurements
to be discussed in Chapter 6. These experiments
deduce the shear viscosity from the damping of collective (breathing) modes in the trapped gas.
A great deal of interest has focused on viscosity experiments because they seem to
reflect ``perfect fluidity", as is expected in many strongly interacting systems, such as
quark-gluon plasmas.
One can ask about the counterpart of perfect fluidity in condensed matter
superconducting systems.
Indeed the conductivity at any low frequency $\omega$ except strictly zero is a close
analogue to viscosity. This conductivity is associated with the excited states of
the condensate and must be distinguished from the infinite conductivity of
the condensate itself which occurs at $\omega = 0$. Despite the fact
that the condensate contribution to conductivity is infinite, (while the condensate
contribution to viscosity is zero), the analogies hold between the contributions
to transport from the condensate \textit{excitations}.
A perfect fluid has anomalously low viscosity, while a bad metal has
anomalously low conductivity. This
``bad metallicity"
is widely studied in the high temperature superconductors
\cite{Emery}.

Recent viscosity data are plotted in Chapter 6 for a unitary trapped
Fermi gas \cite{ThomasViscosityScience}. The viscosity and its ratio to entropy density are both observed to be strongly suppressed at low T
as was theoretically predicted \cite{OurViscosity}.
The normal state behavior is in contrast to what is expected of
a Fermi liquid. The superfluid behavior is more similar
to (fermionic) helium-3 but in contrast to
what would be expected in superfluid (bosonic) helium-4.
The behavior of both helium-3 and the Fermi gases
can be understood as reflecting the strong
reduction in the number of \textit{fermionic}
carriers in the presence of a pairing gap. Because
of the pseudogap,  the normal state is very different from
a Fermi liquid, where a low $T$ upturn would otherwise
be expected. Because the carrier number is effectively
constant in a Fermi liquid, this upturn would derive from
the decrease in inter-particle scattering.
The behavior of the shear viscosity in superfluid helium-4 also shows a low
$T$ upturn. This is thought to reflect the phononic excitations
which dominate the transport in this regime.
In a BCS-like
superfluid the collective mode phonon-like
excitations are longitudinal. In a charged superconductor,
as is well known, they do not
affect the
analogous near-zero frequency
conductivity.
Similarly, to the extent that the unitary Fermi gas has
BCS-like characteristics, one would not expect the phonons
to affect a transverse probe, like the shear viscosity.
This may then explain the measured behavior of the viscosity.

\subsection{Two Photon Bragg Scattering: Analogies with
Neutron Scattering Experiments}

\begin{figure}[thb]
\centerline{\includegraphics[width=2.0in,clip]
{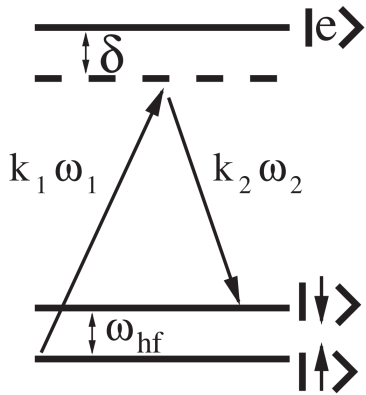}}
 \caption{Experimental Scheme for Bragg Scattering. This represents
a ``spin flip" process in the sense that the final state and the initial
state are different. One can also contemplate non-spin flip processes
in which the initial and final states correspond to the same quantum
index.}
 \label{fig:Bragg}
\end{figure}

The analogue to neutron experiments which have been so important in condensed matter,
are two photon Bragg scattering studies.
A strength of cold atom
experiments is that, unlike neutron probes, they can, in principle,
separately measure the density
and spin correlation functions.
Bragg scattering
can be thought of as a
coherent scattering process involving absorption of a photon
from one of two laser (Bragg) beams and stimulated emission into
the other: a two-photon transition.
This combination of processes can leave the
atoms in the same internal state, but with a new momentum,
k, determined by the geometry and wavelength of the Bragg
beams.
The two lasers have slightly different frequencies, to additionally
account for a shift in energy of the final state.
One can also contemplate processes where the internal states
are changed such as in Figure \ref{fig:Bragg}.
In linear response theory these experiments
correspond to measuring the dynamical density-density or spin density-spin density
correlation functions. These are functions of the momentum and frequency transfer.
Typically, the Bragg response is measured by looking at
the cloud using time-of-flight atom imaging, where the gas is
suddenly released from the trap and allowed to expand before
imaging.

An important aspect of the density-density scattering processes is that
they reflect the phononic-like collective modes of the superfluid order
parameter.  This is in contrast, say, to what is referred to
as ``transverse" transport probes such as the shear viscosity or the conductivity.
In this way, these measurements have the capability of indicating superfluid coherence \cite{OurBraggPRL}.

\subsection{Fermi Gases with Imbalanced Spin Populations}

Unlike conventional superconductors, where spin-polarization is excluded by the Meissner effect, spin imbalance is a readily tunable parameter in ultracold atomic Fermi gases. Shortly after the development of the BCS theory, theorists began to speculate how pairing is modified by spin-polarization. They predicted new and exotic pairing mechanisms, such as the elusive FFLO-state (named after its proposers: Fulde, Ferrell, Larkin, and Ovchinnikov), that may occur in an imbalanced system.  The FFLO-state features pairs with a momentum equal to the difference of the Fermi momenta of the two spin-states.  This non-zero center-of-mass momentum results in an order parameter that is both anisotropic and oscillatory in space.

While spin-polarization is excluded in conventional superconductors, certain compounds such as the heavy-fermion materials, support coexisting superconducting and magnetic order.  While there is some preliminary evidence for FFLO-pairing in these systems, the smoking gun, non-zero pair momentum, has not been found.  Arbitrarily spin-polarized atomic Fermi gases may be created by simply adjusting the relative populations of the hyperfine sublevels corresponding to ``spin-up" and and ``spin-down".  The stability of these states over the time scale (seconds) of the experiments ensures that the spin-polarization of the gas does not vary during the experiment.  Several experiments have been performed in this way with the result that the gas phase separates into an unpolarized superfluid core surrounded by a polarized normal shell \cite{Zwierlein06, Partridge06, Nascimbene09}.  No evidence for the FFLO state has yet been found.

\begin{figure}[tb]
\centerline{\includegraphics[scale=.75,bb=8 11 246 204,clip=true]{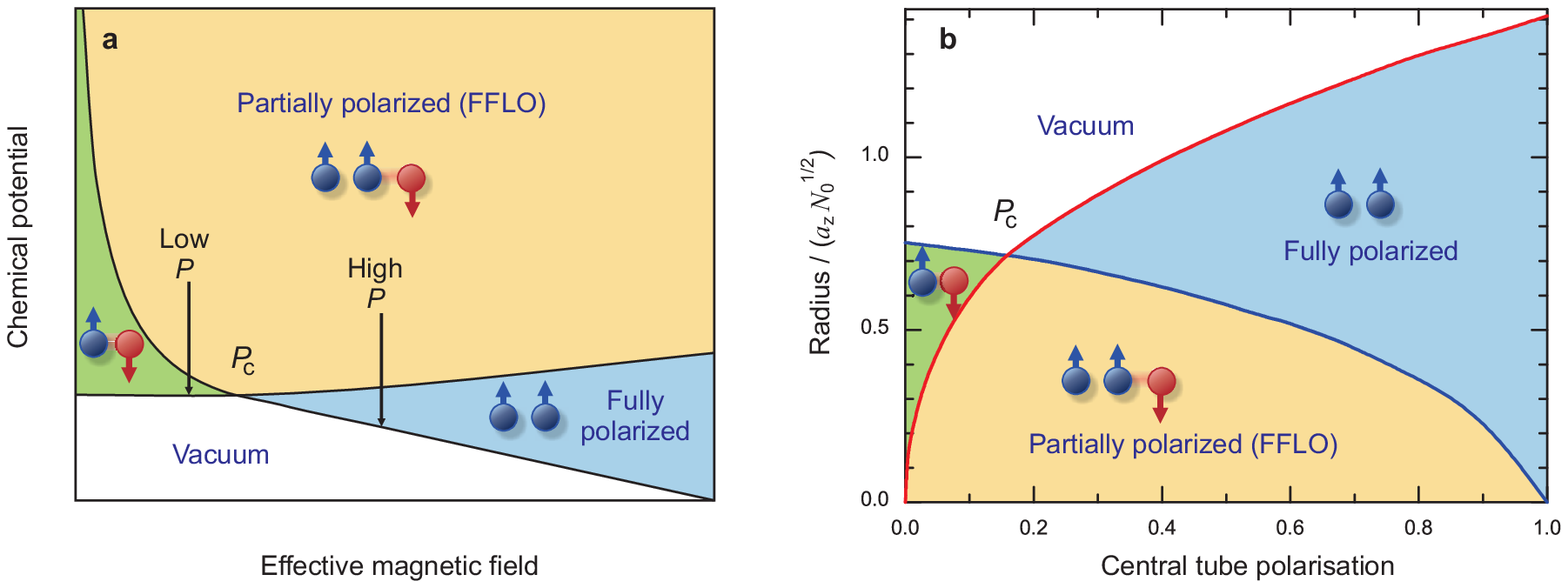}}
\caption{Phase diagram of a one-dimensional spin-imbalanced Fermi gas.  The horizontal axis
is the effective magnetic field $h$, which is related to the degree of polarization.  The vertical
axis is the total chemical potential $\mu$, which is related to the density.  The vertical lines show how the phase diagram is traversed in going from the center of the 1D tube, where $\mu$) is high, to the edges of the tube.  At low imbalance the edge corresponds to a fully paired phase, while for high imbalance the edge is fully polarized.  In both cases, the center of the tube is partially polarized and predicted to the exotic FFLO (Fulde-Ferrell-Larkin-Ovchinnikov) paired state.
(Reprinted from Ref.~\cite{Liao10} and adapted from Ref.~\cite{Orso07})}.
\label{fig:1Dphasediagram}
\end{figure}

 The FFLO state in three dimensions (3D) remains elusive, but theory shows that it is ubiquitous in a spin-imbalanced Fermi gas in 1D \cite{Orso07}.  The phase diagram, shown in Fig.~\ref{fig:1Dphasediagram}, is predicted to have three distinct phases:  1) fully-polarized normal; 2) fully-paired superfluid; and 3) partially-polarized FFLO superfluid.  This result was explored experimentally by using a two-dimensional optical lattice to produce an array of 1D tubes \cite{Liao10}.  By making the lattice intensity sufficiently strong, the tubes are effectively isolated from one another.  Since the atoms are confined harmonically along the tube axis, the density and hence the total chemical potential $\mu = \mu_\uparrow + \mu_\downarrow$ in each tube varies along this axis.  While $\mu$ varies from a maximum value $\mu_0$ at the center of the tube to 0 at the edges, the chemical potential difference $h = \mu_\uparrow - \mu_\downarrow$ must be constant in order for the gas to be in chemical equilibrium.  Traversing the tube from center to the edge corresponds to a cut through the phase diagram, as indicated by the vertical lines in Fig.~\ref{fig:1Dphasediagram}.  If $\mu_0$ is sufficiently large for the center of the tube to be partially polarized, a phase boundary will be encountered in passing from the center of the tube to the edge.  For each combination of $\mu_0$ and $h$, therefore, the tube will exhibit a pair of phases, with the center always being partially polarized while the wings will be either fully paired for small $h$, or fully polarized for large $h$.

\begin{figure}[tb]
\centerline{\includegraphics[width=4.5in,bb=11 17 199 93]{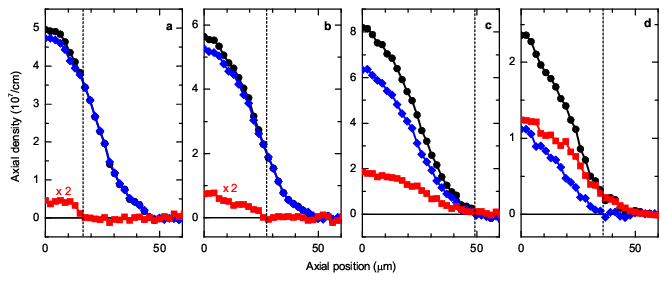}}
\caption{Axial density profiles of a spin-imbalanced 1D ensemble of tubes.
Black circles represent the 1D density of the
majority atoms (spin-up) , the blue diamonds represent the minority (spin-down), while the red squares show the density difference. The dashed vertical lines indicate phase boundaries, where either the minority density or the density difference vanish.  The polarizations $P$ in the central tube are (a) $P=0.015$, (b) $P=0.055$, (c) $P=0.10$, and (d) $P=0.33$.  For low $P$ (a and b), the edge of the cloud is fully paired and the difference is 0, while for larger $P$ (d), the edge is fully polarized and the minority density vanishes.
(Reprinted from Ref.~\cite{Liao10})}.
\label{fig:1Ddensities}
\end{figure}

Each tube was loaded with $\sim$200 $^6$Li atoms with a particular imbalance.  Phase boundaries were extracted from the density distributions obtained from optical imaging.  The fully polarized/partially polarized boundary is determined by where the minority density $n_\downarrow$ goes to zero, while the fully paired/partially polarized boundary is given by the location of vanishing spin density $n_\uparrow - n_\downarrow$.  Several representative density distributions showing the phase boundaries are given in Fig.~\ref{fig:1Ddensities}.  The phase diagram was mapped out experimentally in this way, and good agreement was found with Bethe-ansatz theory \cite{Liao10}.

\subsection{Rotating Gases and Analogue of Magnetic Field Effects}

In contrast to a normal fluid such as water, a superfluid can only rotate
by forming a regular array of quantized vortices, each of which
carries part of the total angular momentum of the superfluid. In addition
to expelling atoms from their centers to leave a string-like hollow core, the
vortices also repel each other to form a regular lattice
pattern.
In trapped atomic gases, this rotation can be created by using
a repulsive ``spoon" potential, created by a blue-detuned laser beam, to vigorously stir the gas.

Charged fermionic superconductors and rotating superfluids are closely related by the
correspondence
\meq{e\bs{A}(\bs{r}) \leftrightarrow m\bs{\omega}\times\bs{r} \label{Aomega}}
between the superconductor with a magnetic field and the neutral rotating
(at frequency $\omega$) superfluid.
It can be said that neutral superfluids are more analogous to extreme
type II superconductors, where the penetration depth is infinite.
Just as in charged superconductors, at sufficiently high rotation frequencies,
the neutral superfluids exhibit vortices (as in Figure 8) and these, in turn, exhibit
quantized circulation.

Experiments to explore vortex phases and possible quantum Hall
phenomena are underway.
Some progress towards pursuing analogies with condensed matter
relates to the precursor diamagnetism observed, say, in
the high $T_c$ cuprates. This diamagnetism
has been the topic of considerable excitement \cite{Ong2}
and much debate. Analogous to this diamagnetism in a charged superconductor
may be an above-$T_c$
reduction in the moment of inertia of a Fermi superfluid. This
is associated with finite size clouds. There may be
some experimental indications
that the equilibrium values of this moment of inertia are somewhat
suppressed in the normal state
\cite{RiedlGrimm}.
Besides high $T_c$ cuprates, this
general class of experiments bears on other important condensed matter
systems
such as possible observation of supersolid phases.

\section{Conclusions}

The aim of this Chapter is to convey
new and exciting developments in the physics of ultracold atoms to
condensed matter physicists.
It should be clear that, because of the shared Fermi statistics,
atomic Fermi gases (and their optical lattice counterparts) have the
potential for addressing important unsolved problems relating to electrons in condensed matter.
They would seem even more promising in this regard than atomic Bose gases.

Another exciting aspect of the Fermi gases is their potential to explore a
generalized form of fermionic superfluidity which seems, in many ways, more natural
than simple BCS theory. In this generalized form, known as BCS-BEC crossover theory,
the pair size must no longer be
large or the pair binding weak. As a result, pairs form at a higher temperature
(called $T^*$ in the literature) than that at which they condense, $T_c$.
BCS theory as originally postulated can
be viewed as a paradigm among theories of condensed matter systems; it
is generic and model independent, and extremely well
verified experimentally for conventional (presumably long coherence length)
superconductors.  The observation that a BCS-like approach
extends beyond strict BCS theory, suggests that there is
a larger theory to be discovered. Equally exciting is the
possibility that this discovery phase can proceed
in a very collaborative fashion, involving both theory and experiment.

One fascinating aspect feature of this crossover
is that the statistics may be tuned continuously from fermionic to bosonic.
This may lead to a fundamental challenge for theory. In any attempt to
combine bosonic and fermionic mean field theories, as must be
accomplished in
BCS-BEC crossover, one should be aware that there is no fully satisfactory
mean field theory of the weakly interacting Bose gas (or Bogoliubov
theory) which addresses the entire range of temperatures.
This is in contrast to the situation
for the fermionic (BCS-based) superfluids which apply to all
temperatures.
The central problem is that extensions of Bogoliubov theory
to the transition region generally lead to a first order transition.
This suggests that either one must use a far more
sophisticated model (than Bogoliubov theory) or a
simpler BCS-based-level theory (in which the bosons are
essentially non-interacting) to avoid
discontinuities at $T_c$, (for example as shown in the inset to Figure 12).

The BCS-BEC crossover picture has been investigated for many years in the context of high temperature superconductors \cite{ourreview}. It leads to a particular interpretation of a fascinating, but not well
understood non-superfluid phase, known as the ``pseudogap state."  In the cold atom
system, this crossover description is not just a scenario, but has been
realized in the laboratory. A number
of cold gas experiments have been interpreted as evidence for a
pseudogap state. Where there seems to be controversy about this
claim is in inferences drawn from thermodynamics \cite{SalomonFL}.
We note, finally, that research in this field has not been
limited exclusively to the two communities (condensed matter and AMO). One has
seen the application of these crossover ideas
and, in particular a focus on the unitary gas, to nuclear physics and
to particle physics as well.
There are not many problems in physics which
have as great an overlap with different subfield communities.

We have tried in this Chapter to stress the powerful tunability of the ultracold
Fermi gases arising from Feshbach and other experimental ``knobs." We have also emphasized the wide-ranging experimental tools which have
been developed by the AMO community over a relatively short period of time.
With these tools and others awaiting development, the future is wide open.

\section*{Acknowledgements}
{RGH acknowledges support from ARO Grant No. W911NF-07-1-0464 with funds from the DARPA OLE program, ONR, NSF, and the Welch Foundation (Grant No. C-1133). KL acknowledges support from
NSF-MRSEC Grant
0820054. We thank John Thomas, Sylvain Nascimbene, Christophe Salomon, Munekazu
Horikoshi and Hui Hu for sharing their data and calculations with us.
We thank A. Fetter, D. Stamper-Kurn, and D. Wulin for useful conversations and their
help with the manuscript.}

\bibliographystyle{Nature}

\bibliography{Hulet,Review2}

\end{document}